\documentclass[12pt]{article}
\usepackage{amssymb}
\usepackage{epsfig}
\usepackage{float}
\usepackage{slashed}
\newcommand{\be}{\begin{equation}}
\newcommand{\ee}{\end{equation}}
\newcommand{\bea}{\begin{eqnarray}}
\newcommand{\eea}{\end{eqnarray}}

\begin{document}

\vskip 1cm

\begin{center}
 \textbf{THE MAXIMAL MASS MODEL AND LEPTON MAGNETIC MOMENT}
\end{center}
\vskip 1cm\begin{center} V. N. Rodionov
\end{center}

\begin{center}
{\em  Moscow State Geological Prospecting University, 118617
Moscow, Russia.}
\end{center}
\begin{center}
{\em vnrodionov@mtu-net.ru}
\end{center}

\begin{center}

\abstract{The two-component formulation of quantum electrodynamics
with fundamental mass is studied.  We review and update the
prediction of the primordial existence of lepton magnetic moment
in framework of two component formulation of the Maximal Mass
Model containing a limiting mass $M$, which is considered as new
universal physical constant. As it well known in the Dirac theory
so-called gyromagnetic factor $g=2$. Quantum electrodynamics
together with electro-weak theory and hadronic contributions
predicts deviations from Dirac's value. It is very important that
all these effects \textbf{slightly increase the $g$-value}. But in
our model we have any decreasing of this quantity
$g=2\sqrt{1-m^2/M^2}$, where $m$ - is a lepton mass. The most
intriguing prediction of new approach is the absolute value of
this deviation increases with growth of a lepton mass. In this
connection the direct experimental measurements of $\tau$ - lepton
anomalous magnetic moment $a_\tau = (g - 2)/2$ gain in
extraordinary importance. The most stringent limit $-0.052 <
a_{\tau} < 0.013$ at 95\% confidence level, was set by the {\small
DELPHI}\cite{delphi} collaboration. The authors also quote their
result in the form of central value and error $ a_\tau = -
0.018(17).$ To pay attention that the sign of $a_\tau$ is the
negative and now we can speak about a qualitative agreement with
our predictions.}
\end{center}

 \vskip 1cm
\begin{center}
 \textbf{1 Introduction.}
\end{center}
\vskip 1cm

Numerous precision tests of the Standard Model ({\small SM}) and
searches for its possible violation have been performed in the
last few decades, serving as an invaluable tool to test the
contemporary quantum field theory (QFT). The very concept of
elementary particle assumes that it does not have a composite
structure. In agreement with the recent experimental data such a
structure has not been disclosed for no one of the fundamental
particles of the SM, up to distances of the order of $10^{-17} -
10^{-18}$ cm. The adequate mathematical images of point like
particles are the local quantized fields - boson and spinor.
Intuitively it is clear that the elementary particle should carry
small enough portions of different "charges" and "spins". In the
theory this is guaranteed by assigning the local fields to the
lowest  representations of the corresponding groups.

As for the mass of the particle $m$, this quantity is the Casimir
operator of the \emph{\textbf{noncompact}}  Poincar\'{e} group and
in the unitary representations of this group, used in QFT, they
may have arbitrary values in the interval $0 \leq m <\infty$. In
the SM one observe a great variety in the mass values. For
example, t-quark is more than 300000 times heavier than the
electron. In this situation  the question naturally arises: up to
what values of mass one may apply the concept of a local quantum
field? Formally the contemporary QFT remains logically perfect
scheme and its mathematical structure does not change at all up to
arbitrary large values of quanta's masses.

In 1965 M. A. Markov  \cite{Markov} pioneered the hypotheses
according to which the mass spectrum of the elementary particles
should be cut off at the Planck mass $m_{Planck} = 10^{19} GeV $ :
\begin{equation}\label{mpl}
 m \leq m_{Planck}.
\end{equation}
The particles with the limiting mass $m = m_{Planck}$, named by
the author "maximons"  should  play special role in the world of
elementary particles. However, Markov's original condition
(\ref{mpl}) was purely phenomenological and he used standard field
theoretical techniques even for describing the maximon.

Till recently one can see no reason why SM should not be adequate
up to value of order the Planck mass. But we are living in times,
where many of the basic principles of physics are being challenged
by need to go beyond SM. By now it is confirmed that dark matter
exists and it consists of a large fraction of the energy density
of the universe ($\sim$ 25 percent) \cite{dark} while dark energy
consists of $\sim$ 70 percent. The energy density of the
non-baryonic dark matter in the universe is known to be
\cite{dens} \bea \Omega_{DM}h^2 = 0.112 \pm 0.009 \label{odark}
\eea where $\Omega_{DM}$ is the energy density in units of the
critical density and $h\sim 0.71$ is the normalized Hubble
parameter. Since the visible matter consists of only $\sim$ 5
percent of the matter of the universe, the laws of physics or laws
of gravity as we know today may not be sufficient to explain the
dark matter and dark energy content of the universe.

In this connection
 a more radical approach was developed \cite{Kad} - \cite{KMRS}.
The Markov' s idea about existence of a maximal value for the
masses of the elementary particles has been understood as a new
fundamental principle of Nature, which similarly to the
relativistic and quantum postulates should be put in the grounds
of QFT. Doing this the condition of finiteness of the mass
spectrum should be introduced by the relation:
\begin{equation}\label{Mfund}
m \leq  M,
\end{equation}
where the maximal mass parameter M called the
"\emph{\textbf{fundamental mass}}" is a \emph{\textbf{new
universal physical constant}}.  Now objects for which $ m
> M$ cannot be considered as elementary particles, as  to them does not
correspond a local field.

A \emph{\textbf{new concept}} of a local quantum field has been
developed on the ground of (\ref{Mfund}) an on simple geometric
arguments, the corresponding Lagrangians were constructed  and an
adequate formulation of the principle of local gauge invariance
has been found. It has been also demonstrated that the fundamental
mass $M$ in the new approach plays the role of an independent
universal scale in the region of ultra high energies $E \geq M $.

The above-presented approach allows a simple geometric realization
if one considers that the fundamental mass $M$ is the curvature
radius of the momentum anti de Sitter 4-space ($\hbar = c =1$)
\begin{equation}\label{O32}
    p_0^2 - p_1^2 - p_2^2 - p_3^2 + p_5 ^2 = M^2.
\end{equation}
For a free particle, for which  $p_0^2 - \overrightarrow{p}^2 =
m^2$, the condition (\ref{Mfund})  is automatically satisfied on
the surface (\ref{O32}). In the approximation
\begin{equation}\label{Plpred}
    |p_0|,\;\;|\overrightarrow{p}| \ll M, p_5 \cong M.
\end{equation}

\noindent the anti de Sitter geometry does not differ from the
Minkowski geometry in four dimensional pseudo--Euclidean $p$-space
("flat limit").

However, it is much less obvious that in  the momentum 4-space
(\ref{O32}) one may fully develop the apparatus of quantum field
theory, which after transition to configuration representation
(with the help of a specific 5-dimensional Fourier transform)
looks like a local field theoretical formalism in the four
dimensional $x$-space \cite{Kad}, \cite{KM}. It is fundamentally
important that the new theory may be formulated in a gauge
invariant way \cite{Kad} - \cite{KMNC}. In other words, in the
considered \emph{\textbf{geometric}} approach there are conditions
to construct an adequate generalization of the Standard Model,
which were called the Maximal Mass Model \cite{Max}.

In the new approach the electromagnetic potential becomes a
5-vector associated with the corresponding de Sitter group. The
extra fifth component does not connect with the independent
dynamical degree of freedom. The gauge invariant equation of
motion, replacing the Dirac-Maxwell equations, are set up in the
framework of an appropriate Lagrangian formalism.

Note one of the most interesting consequence of this approach that
the new formulation of the electromagnetic interactions is minimal
with respect to the 5-potential but is not so in terms of the
usual 4-potential. As a result, the underlying physics looks much
richer than the ordinary electromagnetic phenomena.
 In particular the new scheme predicts the primordial existence of the
 so-called the anomalous magnetic moment
(AMM) of particles (the dimensionless quantity $a$ is defined as
$a=(g-2)/2$, where g is gyromagnetic factor).  In spite of the
fact that the value of $M$ may be high enough one should bear in
mined that the recent experiments on measurement of electron and
muon $(g-2)$-factors are reached the fabulous relative precision
~\cite{odom}, \cite{bnl}. Another experimental developments allow
to significantly  improve the prediction for AMM of the $\tau$
lepton \cite{edel}.

In 1958, Feynman and Gell-Mann~\cite{FG} revived the two-component
fermion theory in the context of their work on the V-A form of the
weak interactions. Then this was studied extensively by many
authors, who reformulated quantum electrodynamics in the form of
two-component theory. In this article we review and update the
prediction of the primordial existence of lepton magnetic moment
in framework of two component formulation of the Maximal Mass
Model.

\begin{center}
 \textbf{2 The two-component equations of motion for de Sitter fermion fields }
\end{center}

Let us apply the developed in \cite{Kad} - \cite{KMRS} methods to
describe the interaction between a neutral abelian vector field
and a
 charged fermion field (with charge $e$ and mass $m$).
 Remember that in the new scheme the electromagnetic potential,
similarly to the momentum, becomes a 5-vector $ A_L(x, x_5) = (
A_\mu(x, x_5),A_5(x, x_5) )$.
 Thus we have the new  description is
based on the gauge invariant set equations (see, for instance,
~\cite{Kad}) for the 4-component wave function $\Psi(x,x_{5})$:

\begin{equation}\label{Ltot}
\Big(i\widetilde{D}_L \gamma^L +M \gamma_5-2M\sin
\mu/2\Big)\Psi(x,x_5)=0;\,\,\,\, \gamma^L =
(\gamma^0,\gamma^1,\gamma^2,\gamma^3,\gamma^5);
\end{equation}
\begin{equation}\label{L}
 \Big(
\widetilde{D}_L \widetilde{D}^L + M^2
\Big)\Psi(x,x_5)=0;\,\,\,\,\sin\mu =m/M;\,\,\,\,\, L=0,1,2,3,5,
\end{equation}
 in which all components  of the 5-gradient $\widetilde{D}_L= (\widetilde{D}_\mu,\widetilde{D}_5) $
  is defined as
 the covariant derivatives
\begin{equation}
    \begin{array}{c}
     \widetilde{D}_\mu = \partial_\mu + i e\, A_\mu(x, x_5)e^{-i{M}x_5};  \\
      \widetilde{D}_5 = \partial_5 + i e\, A_5(x, x_5)e^{-i{M}x_5}. \\
    \end{array}
\end{equation}

The equation of the first order (\ref{Ltot})\footnote {Which one
may consider as new Dirac equation.}  can be transformed into the
equation of the second order by application to it the operator
\begin{equation}\label{24}
\widehat{D}=i\widetilde{D}_L \gamma^L +M \gamma_5 + 2M\sin \mu/2.
\end{equation}
Thus  we have the following equation of a fermion motion in
5-dimensional space

$$
\Big[i\widetilde{D}_L \gamma^L +M \gamma_5+2M\sin \mu/2\Big]\Big[
i\widetilde{D}_L \gamma^L +M \gamma_5-2M\sin
\mu/2\Big]\Psi(x,x_5)=
$$
\begin{equation}\label{25}
=  \Big[-\widetilde{D}_L \widetilde{D}^L- M^2  +2M \Big (M \cos
\mu -i\frac{\partial}{\partial x_5}\Big)-\frac{i }{2}e F_{L M}
\Sigma^{L M }\Big]\Psi(x,x_5) = 0,
\end{equation}where
$ \Sigma^{L M}=\frac{1}{2}\big( \gamma^L \gamma^M-\gamma^M
\gamma^L \big)$;  and $F_{L M}$ - is the 5-dimensional field
strength
\begin{equation}
F_{K L}(x,x_5)=\frac{\partial}{\partial x^L}\Big[ e^{i x_5 M}
A_K\Big]-\frac{\partial}{\partial x^K}\Big[ e^{i x_5 M} A_L\Big].
\end{equation}

 Let us emphasize that
in (\ref{25}) all functions is defined in five dimensional
configuration space.
 The equations
(\ref{Ltot})-(\ref{25}) involves also the variables
$\frac{i}{{M}}\frac{\partial A_\mu}{\partial x_5}$ and
$\frac{i}{{M}}\frac{\partial A_5}{\partial x_5}$, which have an
auxiliary character. Moreover, (\ref{Ltot})-(\ref{25}) depends on
the component $A_5$ which is a gauge degree of freedom. If $F_{K
L}$ in (\ref{25}) is fixed, then one can consider it as the motion
equation of charged particle  in an external electromagnetic
field. In this case the appropriate dependence of $F_{K L}$  is of
the form
$$
F_{K L}(x,x_5)= F_{K L}(x,0).
$$
One may easily exclude all odd quantities and using (\ref{L}),
instead eq. (\ref{25}), obtain

\begin{equation}\label{27}
2M\Big( M \cos \mu  -i\frac{\partial }{\partial x_5}- \frac{i }{4
M}e F_{\mu \nu} \sigma^{\mu \nu }\Big)\Psi(x, x_5)  = 0,
\end{equation}
where $\sigma^{\mu \nu}=\frac{1}{2}\big(
\gamma^\mu\gamma^\nu-\gamma^\nu\gamma^\mu \big)$ and $F_{\mu \nu}$
- is the Maxwell field strength, $\mu,\nu = 0,1,2,3$.

Let us proceed now a derivation of 4-sector equation attached to
the physical plane $x_5=0$ \footnote {With
 the details one may be acquainted, for instance, in \cite{Kad}.}.
Thus we project (\ref{27}) onto this plane using eq. (\ref{L}). It
gives

\begin{equation}\label{33}
\Big[ {D_\mu}^2 + m^2 + \frac{i }{2}e F_{\mu \nu} \sigma^{\mu \nu
}\cos\mu\left. + \Big(\frac{1 }{4M}e F_{\mu \nu} \sigma^{\mu \nu
}\Big)^2\Big]\Psi(x,x_5)\right |_{x_5=0}=0,
\end{equation}
where $ D_{\mu} = \partial_\mu + i e A_\mu  $.  Let us consider
the weak field limit, when the field strength is low in comparison
with a so-called fundamental field \cite{Rod}
$$
  F\ll   F^* = \frac{M^2}{e} \simeq 4.41 \cdot 10^{13}
     \frac{M^2}{m^2},
$$
where $m$ - is an electron mass. It is easy to see that under $
M\sim \,1\, TeV $ the fundamental field  is $F^* \sim  10^{20}\,
G$ and with a good accuracy
 we may neglect the last term in (\ref{33}).
Hence in the linear  approximation over the electromagnetic field
instead (\ref{33}) we may obtain

\begin{equation}\label{31}
\Big[{D_\mu}^2 + m^2 + \frac{i }{2}e F_{\mu \nu} \sigma^{\mu \nu }
\cos\mu \Big]\Psi(x,0) = 0.
\end{equation}

Here the last term defines the interaction  of the charged lepton
having magnetic moment $ \widetilde{\mu} =\mu_0 \cos\mu $, where
$\mu_0 =e/2m$ - is the Bohr's magneton, with electromagnetic
field. It is clear that under $M\rightarrow \infty$ we formally
have

$$
\cos \mu =1,
$$
 and therefor new value of a magnetic moment only
 in the "flat limit"
coincides with predicts of Dirac theory of charged point-like
spin-$1/2$ particle. Note that since in (\ref{31}) any term has
either no gamma matrix or two, one may write this equation in
terms of right- and left-handed spinors.

In the case of the constant magnetic field the equation (\ref{31})
takes the form
\begin{equation}\label{34}
\big[ {D_\mu}^2 + m^2  - e \cos\mu \vec{\Sigma}
\vec{H}\big]\Psi=0,
\end{equation}
where
$$
\vec{\Sigma}= \Big(\begin{array}{cc} 0& \vec{\sigma } \\
                              \vec{\sigma}  & 0
                              \end{array}\Big),$$
and $ \overrightarrow{\sigma} $ - is the Pauli matrixes.
 The experiments measure the  gyromagnetic factor
$g$, defined by the relation between the particle's spin $\vec{s}$
and its magnetic moment $\vec{\widetilde{\mu}}$,
\be \vec{\widetilde{\mu}}=g_0 \mu_0 \cos\mu\vec{s}. \ee
 As it well known in
the Dirac theory $g_0=2$. Quantum electrodynamics together with
electro-weak theory and hadronic contributions predicts deviations
from Dirac's value. It is very important that all these  effects
\textbf{slightly increase the $g$-value}. But in our model we have
any decreasing of this quantity
\begin{equation}
\widetilde{g}=2\sqrt{1-m^2/M^2}.
\end{equation}

 The most intriguing prediction of new approach is the absolute
value of this deviation increases with growth of a lepton mass. In
this connection the direct experimental measurements of $\tau$ -
lepton  anomalous magnetic moment gain in extraordinary
importance.

\begin{center}
 \textbf{3 Conclusions}
\end{center}

As it was mentioned conventional to express the difference of $g$
from 2 in terms of the value of AMM, a dimensionless quantity
defined as $a = (g - 2)/2$. In considering approach we  have

\begin{equation}\label{34}
a= \cos\mu-1 = -2 \sin^2\mu/2,
\end{equation}
and for $m \ll M$ our result is

\begin{equation}\label{35}
a(M)=  - \frac{m^2}{2 M^2}.
\end{equation}

The AMM of the electron, $a_e$, is rather insensitive to strong
and weak interactions, hence providing a stringent test of {\small
QED} and leading to the most precise determination of the
fine-structure constant~\cite{Gabrielse_a_2006}. On the other
hand, the $g-2$ of the muon, $a_{\mu}$, allows to test the entire
{\small SM}, as each of its sectors contributes in a significant
way to the total prediction. Compared with $a_e$, $a_{\mu}$ is
also much better suited to unveil or constrain new effects.
Indeed, for a lepton $l$, their contribution to $a_l$ according to
(\ref{35}) is proportional to $m_l^2/M^2$, where $m_l$ is the mass
of the lepton and $M$ is the fundamental mass. Thus leading to an
$(m_{\mu}/m_e)^2 \sim 4\times 10^4$ relative enhancement of the
sensitivity of the muon versus the electron anomalous magnetic
moment. This more than compensates the much higher accuracy with
which the $g$ factor of the latter is known.

It became clear that the accuracy of the theoretical prediction of
the muon $g$$-$$2$, challenged by the E821 experiment underway at
Brookhaven~\cite{bnl}, was going to be restricted by our knowledge
of its hadronic contribution.  This problem has been solved by the
impressive experiments at low-energy $e^+e^-$ colliders, where the
total hadronic cross section (as well as exclusive ones) were
measured with high precision, allowing a significant improvement
of the uncertainty of the leading-order hadronic
contribution~\cite{edel}. According to final report of the muon
E821 AMM measurement at BNL ~\cite{BNL} the difference between the
measured and theoretical values of muon is equal

\begin{equation}\label{361}
\Delta a_\mu = (22 \div 26)\cdot10^{-10}.
\end{equation}
The principal conclusion drawn from a comparison of the above
estimates is that we cannot rule out the possibility that the
observed difference between the theoretical and experimental
values for $\Delta_\mu $ is equal to $|a_\mu (M)|$.  Perhaps the
parameter $ M$ in the new theory is related to the Higgs boson
mass ($M_H$). In this case, tile difference between ${a_\mu}^
{exp}  $ and ${a_\mu}^ {SM}$ can provide valuable information
about the particle whose mass has not been determined in the
standard model. Substituting $m_{\mu}$ and the anomalous magnetic
moment of a muon into (\ref{35}), we can easily impose the
following constants on the $H$-boson mass:
$$
           1.46 \, {\rm TeV}  \le M_H \le 1.58 \, {\rm TeV}.
$$

The anomalous magnetic moment of the $\tau$ lepton, $a_{\tau}$,
would suit even better because $(m_{\tau}/m_e)^2 \sim 1.2\times
10^7$. However, its direct experimental measurement is prevented
by the relatively short lifetime of this lepton, at least at
present. The existing limits are based on the precise measurements
of the total and differential cross sections of the reactions
$e^+e^- \to e^+e^-\tau^+\tau^-$ and $e^+e^- \to Z \to
\tau^+\tau^-\gamma$ at {\small LEP} energies. The most stringent
limit, $-0.052 < a_{\tau} < 0.013$ at 95\% confidence level, was
set by the {\small DELPHI}\cite{delphi} collaboration. The authors
also quote their result in in the form of central value and error:

\begin{equation}\label{36}
a_\tau = - 0.018(17)
\end{equation}
To pay attention that the sign of (\ref{36}) is the negative. It
is clear that the sensitivity of the best existing measurement of
$a_\tau$ is still more than an order of magnitude worse then
needed~\cite{edel} but now we can speak about a qualitative
agreement (\ref{35}) with (\ref{36}). This gives grounds that our
calculation of the particle AMM (\ref{34}) can be considered as a
correct. In this case we can see that  maximon ($m=M$) AMM takes
the value $a_{maximon} =-1$, i.e. the maximon gyromagnetic factor
should exactly equal to zero.

 {\bf Acknowledgements.} The author is very grateful to Prof. V.G.Kady\-shev\-sky,
 provided the inspiration to study the present subject. I am
 sincerely grateful to may colleagues at Dubna, Drs. M.Mateev and
 A.Sorin for their fruitful and benefitting collaboration.
 This work has been supported in part by
 the Program for Supporting Leading
Scientific Schools (Grant No. NSh-3312.2008.2) .

\end{document}